\documentclass{article}
\usepackage{setspace}
\usepackage[utf8]{inputenc}
\usepackage{amsmath}
\usepackage{slashed}
\usepackage{amssymb}
\usepackage{graphicx}
\usepackage{authblk}
\graphicspath{ {./images/} }
\usepackage{comment}
\usepackage{tikz}
\usepackage{geometry}
\usepackage{hyperref}
\hypersetup{ 
colorlinks=true,
linkcolor=blue,
filecolor=magenta,
urlcolor=blue,
citecolor=blue,
pdfpagemode=Fullscreen
}

\usepackage[normalem]{ulem}
\usepackage{enumerate}
\usepackage[shortlabels]{enumitem}

\title{A Fully Solvable Model of Fermionic Interaction in 3+1d}
\author[1]{Seth Grable}
\author[1]{Max Weiner}
\affil[1]{Department of Physics, University of Colorado Boulder, Colorado 80309, USA}
\date{Febuary, 2023}

\begin{document}
\begin{spacing}{1.5}
\maketitle

Recently, Romatschke found that the poles in \(O(N)\) scalar theories do not affect observables such as temperature and pressure. Romatschke went on to show this result holds for marginal, relevant, and irrelevant operators in \(3+1d\) \(O(N)\) scalar theories. We continue in this direction by studying large-\(N\) fermi-interactions in \(3+1d\). To do so, we produce a model of marginally coupled fermi-interactions which is fully renormalizable at large-\(N\). This theory contains poles in the running coupling, however we argue these poles do not affect any physical observables. Further, our theory contains first order phase transition which separates a stable, meta-stable, and unstable phase.

\begin{center}
\item\section{Introduction}
\end{center}

The Gross–Neveu (GN) model is a relativistic field theory of four-fermi interactions. Four-fermi interactions are fundamental and fruitful in: providing models of superconductivity such as in BCS theory, giving an effective, or low energy, theory of nucleons and mesons via the NJL-model, and providing a dual to the quantum Sine-Gordon model via the Thirring model. However, BCS theory is non-relativistic, the Thirring model is nonrenormalizable in \(3+1d\) \cite{banerjee1997explicit}, and the NJL-model is nonrenormalizable in \(3+1d\), not UV complete. Nonetheless, given the successful history of four-fermi interactions, we would like to be able to write down a renormalizable and UV complete theory of such interactions to act as a toy model for QCD in \(3+1d\).

To do so we turn to a large-N Gross-Neveu-inspired model in \(3+1d\). In \(1+1d\) the GN-model is a well-known toy model for QCD exhibiting asymptotic freedom and a negative \(\beta\)-function, and in \(2+1d\) the GN-model mimics planar superconductors. Indeed, in \(1+1d\) the large-\(N\) GN model is likewise asymptotically free and renormalizable \cite{moshe2003quantum}\cite{gracey1990three}\cite{gross1974dynamical}. However, in \(1+1d\) the running coupling contains a pole when approaching the IR, demanding an IR cutoff \cite{moshe2003quantum}. Further, when working in \(3+1d\), four-fermi interactions cannot be renormalizable via a running coupling (in the standard sense). Instead, a term containing a fractional power of the fermionic fields must be removed from the Lagrangian \cite{parisi1975theory}. Further, Moshe et al. \cite{moshe2003quantum} claims the non-renormalizability of four-fermi interactions in \(3+1d\) is due to the algebra of the \(\gamma\)-matrices which causes an infinite number of four-fermion interactions to mix under renormalization. Moshe et al. \cite{moshe2003quantum} also claims the GN model is renormalizable in \(1+1d\) because the symmetry of the model is effectively \(O(N)\). However, we note that the renormalizability of GN in \(1+1d\) can also be seen as an artifact of the theory's marginal coupling. With this, our strategy for renormalization starts by writing down a marginal GN-like model in \(3+1d\) with fractional interactions, which we claim is related to renormalization schemes suggested by Parisi \cite{parisi1975theory}. In doing so, our theory has many features of the GN model in \(1+1d\), such as renormalizability, a negative \(\beta\)-function, and a pole in the IR. Next, we make use of the Ai, Bender, and Sarkar (ABS) conjecture \cite{bender2022pt} by implementing a \(\mathcal{PT}\)-symmetric extension of the theory to ``see around" the pole found in the IR. Where ``see around" the pole means: 1) we can see into the UV using the standard theory, and we can see into the IR using the \(\mathcal{PT}\)-symmetric theory, and 2) poles in the running of the coupling do not affect the theories observables and both theories have identical observables. Points 1) and 2) will be fully illustrated in section \ref{section 3}. Together the strategies of fractional interactions and \(\mathcal{PT}\)-symmetric extensions yield two fully renormalized fermionic theories in \(3+1d\) with identical physics, which together give a complete theory which is UV and IR complete.

A brief discussion is warranted around the topic of large-\(N\) theories with fractional interactions. Indeed, standard perturbation theory is not possible with fractional interaction terms \cite{peskin2018introduction}. However, in large-\(N\) theories \(\frac{1}{N}\) acts as a coupling independent expansion parameter. The coupling independence of \(N\) gives way to renormalizable and solvable theories without the use of traditional perturbation theory \cite{moshe2003quantum}\cite{romatschke2019finite} via the use of saddle point integration \footnote{also know as the method of steepest decent}. In other words, our perturbation method is no longer defined as solving a free theory multiplied by series expansions of the interaction terms. This results in renormalizable and solvable theories with arbitrary interaction terms. Recognition of the \(N\)-field expansion parameter as a way to tackle ``nonrenormalizable" fields was first recognized by Parisi in the 1970s \cite{parisi1975theory}. However, many recent papers have taken advantage of the large-\(N\) expansion parameters to present theories which are analytically solvable for all couplings,  generate transport coefficients, and calculate viscosities
\cite{pinto2020three}\cite{romatschke2019finite}\cite{romatschke2019thermal}\cite{romatschke2021shear}\cite{Weiner:2022vwa}. 
Further, some large-\(N\) theories are both solvable and renormalizable with fractional interactions in arbitrary dimensions \cite{grable2022theremal}. 

Next, we address the strategy of studying \(\mathcal{PT}\)-symmetric versions of related Hermitian theories. It has been known for several decades that \(\mathcal{PT}\)-symmetric extensions of Hermitian theories, which are themselves non-Hermitian, still possess a positive, discrete, and real eigenspectrum \cite{bender1999pt}. However recently Ai, Bender, and Sarkar gave a recipe for formulating \(\mathcal{PT}\)-symmetric field theories from cousin Hermitian theories via analytically continuing the coupling \cite{bender2022pt}:
\begin{equation}\label{eq100}
    \ln Z^{\mathcal{PT}} (g) = \frac{1}{2}\ln Z_{\text{Herm}}(\lambda \rightarrow{-g +i0^+}) + \frac{1}{2}\ln Z_{\text{Herm}}(\lambda \rightarrow{-g -i0^+}).
\end{equation}
Although \cite{bender2022pt} only explicitly covers scalar theories other works \cite{BenderFermi}\cite{Mavromatos:2021hpe} have shown the ability to generate \(\mathcal{PT}\)-symmetric fermionic theories. Particularly \cite{BenderFermi} has shown the existence of \(\mathcal{PT}\)-symmetric GN theory in \(3+1d\). However, we note that the product of \(\Bar{\Psi}\Psi\) is Hermitian, and therefore any product of the form \(g\big(\Bar{\Psi}\Psi\big)^\epsilon\) for any \(\epsilon\in \mathbb{R}\) is also Hermitian and is likewise \(\mathcal{PT}\)-symmetric. If we look at theory with interaction terms \(-g\big(\Bar{\Psi}\Psi\big)^\epsilon\) the problem is not that the Lagrangian is non-Hermitian, but that it is not asymptotically free or bounded from below. Indeed, the fact that the Lagrangian \(L(g)\) \textit{is} asymptotically free and maps to the local limit of a Yukawa model is Gross and Neveu's original motivation for defining the coupling to be positive \cite{gross1974dynamical}. However, in recent work,  Ai, Bender, and Sarkar conjecture that \(\mathcal{PT}\)-symmetric theories which inherit ``the wrong coupling sign" are indeed well defined via analytic continuation. Further works that relate \(\mathcal{PT}\)-symmetric theories to traditional Hermitian theories include refs:
\cite{mavromatos2023schwinger}
\cite{chernodub2021spontaneous} \cite{Mavromatos:2021hpe} \cite{mannheim2019goldstone} \cite{fring2020goldstone}  \cite{fring2020pseudo} \cite{fring2022massive} and \cite{alexandre2018spontaneous}. Work by \cite{alexandre2018spontaneous} demonstrates  Goldstone's theorem for \(\mathcal{PT}\)-symmetric theories and that by \cite{mavromatos2023schwinger} demonstrates RG UV/IR flows between Hermitian and non-Hermitian in the context of Yukawa interactions which in fact have been interpreted in \cite{Mavromatos:2021hpe} as leading to the dynamical appearance of \(\mathcal{PT}\)-phases. Both conclusions support this work as we do locate a Goldstone mode and a first order phase transition.

Further, ref \cite{alexandre2018spontaneous} illustrates that a \(\mathcal{PT}\)-symmetric Lagrangian of a two component scalar field theory results in a pair of inconsistent Euler-Lagrange equations. Nonetheless \cite{alexandre2018spontaneous} gets past this dilemma by claiming there must be non-vanishing surface terms which show up when extremizing the action of a \(\mathcal{PT}\)-symmetric theory. However, Mannheim \cite{mannheim2019goldstone} proposes an alternative solution to the problem of inconsistent Euler-Lagrange equations for \(\mathcal{PT}\)-symmetric theories. Specifically for theories containing two complex scalar fields Mannheim \cite{mannheim2019goldstone} shows that a Hermitian and a \(\mathcal{PT}\)-symmetric version of the theory can be related by a non-unitary similarity transformation. Further, this similarity transformation preserves equal-time canonical commutation relations of the Hermitian theory, i.e. the physical content of the Hermitian and \(\mathcal{PT}\)-symmetric theory are identical. Refs \cite{fring2020goldstone}, \cite{fring2020pseudo}, and \cite{fring2022massive} continue in the direction of Mannheim in providing similarity transformations from Hermitian to non-Hermitian theories in order to generalize Goldstone's theorem for non-Hermitian theories.
  
Our Paper first leverages the ABS conjecture to write down two related theories with opposite sign couplings using equation \eqref{eq100}. Next, we show both theories contain the same physics. So, although there is no simple way (at least known to these authors) to write down a similarity transformation relating our two N-component theories, in the spirit of Mannheim, we claim the \(\mathcal{PT}\)-symmetric extension of our Hermitian theory is well defined. Thus, this paper is not dependent on whether or not the ABS conjecture holds, but was rather motivated by its conjecture to build related theories. Further, in earlier well cited work \cite{bender1998real} Bender notices the novel fact that ``Replacing conventional \(g\phi^4\) or \(g\phi^3\) theories by \(-g\phi^4\) or \(ig\phi^3\) theories has the effect of reversing signs in the beta function." In total this paper puts these ideas in tandem as follows: 

\begin{enumerate}
    \item We write down a marginal Hermitian theory of fermionic interactions in \(3+1d\).
    \item We show this theory is related to a theory with opposite sign coupling by having identical physics at exact large-\(N\), and by having an opposing sign beta-function.
    \item We can then use either beta-function to analyze either theory as they are physically identical. 
\end{enumerate}

This strategy follows the recent work of Romatschke \cite{Romatschke2022} in which he takes advantage of the novel fact that flipping the sign of the coupling also flips the sign of the \(\beta\)-function (as noted by Bender \cite{bender1998real}). The newly found asymptotic freedom is then used to see into the UV. More explicitly: \cite{Romatschke2022} shows the \(\beta\)-function of the standard \(\phi^4\)-theory ends at a pole and the \(\beta\)-function of the cousin \(\mathcal{PT}\)-symmetric theory begins at that same pole and then exhibits asymptotic freedom into the UV. The catch is that both theories generate the same thermodynamics, and therefore one can now follow the running coupling through the pole into UV with the use of \(\mathcal{PT}\)-symmetric theory. 

With all of this in mind, we further note the fact that large-\(N\) theories with opposite sign couplings contain the same physics could either follow from the ABS conjecture, or could follow from the existence of some non-unitary similarity transformation acting on a large-\(N\) theory. However, it can also be seen as simple artifacts of large-\(N\) theories. That being said, at exact large-\(N\), a running coupling renormalization scheme absorbs any overall constants or signs in front of the coupling, leaving the theory invariant to such changes. Thus renormalization at large-\(N\) demands that any differences between related\footnote{By related theories we mean theories that are identical up to an arbitrary complex constant in front of the interaction terms}
theories shows up explicitly in the beta-functions, and not in the physical observable. Put more boldly, Large-N theories are immune to the sickness of Hamiltonians which are unbounded from below.

\begin{center}
\item\section{A Large-$N$ Fermionic Model}
\end{center}

Consider a marginal \(U(N)\) Fermionic theory in \(d\)-dimensions where \(\psi_f = (\psi_1, \psi_2 ... \psi_N\)) in the limit \(N\rightarrow \infty\),  with the Euclidean action 
\begin{equation}\label{eq1}
 A^{(d)} = \int d^dx \bigg[\Bar{\psi}_f\left(\slashed{\partial}+m_b\right) \psi_f + \bigg(\frac{\alpha}{N}\bigg)^\frac{1}{d-1}\big(\Bar{\psi}_f\psi_f )^{\frac{d}{d-1}}\bigg],
\end{equation}
where \(\psi_f(x)\) with \(f = 1,2,...N\) are  \(N\times 4\)-component Dirac spinors, and \(m_b\) is the bare mass. The sign convention of the coupling follows \cite{diab2016and} and \cite{parisi1975theory}, and allows us to smoothly connect to a Yukawa-Gross-Neveu model in \(d=4-\epsilon\) dimensions. To calculate the partition function \(Z^{(d)}(T) = \int \mathcal{D} \Bar{\psi}\mathcal{D}\psi e^{-A^{(d)}}\) of an \(U(N)\) model we multiply by a convenient factor of 1 \cite{grable2022theremal}, 
\begin{equation}\label{eq2}
    1= \int\mathcal{D}\sigma\mathcal{D}\xi e^{\int_x\int_0^\beta i\xi(\sigma - \frac{\Bar{\psi}_f\psi_f}{N})},
\end{equation}
and let the Euclidean dimension of time be compactified to a thermal circle of radius \(\beta\) \cite{laine2016basics}.
For simplicity we let \( \xi \rightarrow iN\times \xi\), and \(\sigma \rightarrow \frac{\sigma}{\alpha}\) to evenly distribute the coupling in \(Z(T)\). Next, we take saddle point integrals around the mean-field values of \(\xi\) and \(\sigma\) (\({\xi_0}\) and \(\sigma_0\)), which are exact in the large N limit. Applying the saddle point condition for the \(\sigma\) field and keeping only the positive solution in terms of \(\xi_0\) (which is equivalent to keeping only positive effective mass terms) \(Z(T)\) is given as,
\begin{equation}\label{eq3}
\begin{split}
 Z^{(d)}(T) = \int \mathcal{D}\bar{\psi} \mathcal{D}{\psi} d\xi_0 \exp \bigg[-N\int_x dx \Bigg( \Bar{\psi} \big(\slashed{\partial} +m_b + \xi_0 \big)\psi - \frac{\xi_0^{d}}{ \alpha}\bigg(\frac{1}{d}\bigg)\bigg(\frac{d-1}{d}\bigg)^{d-1}\Bigg)\bigg].
\end{split}
\end{equation}
Since \(\xi_0\) is a constant, the integral over \(\xi_0\) can be found via saddle point integration and is exact in the large-\(N\) limit. Further, at finite temperature, the \(\slashed \partial\)-operator can be diagonalized in terms of its fermionic Matsubara frequencies \(\omega^f_n = 2\pi T (n+\frac{1}{2})\)  with \(n \in \mathbb{Z}\). After diagonalizing \(\slashed \partial\) and integrating \(\xi_0\) we let \(\xi_0\rightarrow m\), in noting \(\xi_0\) acts as the effective mass. Up to an overall constant, the partition function is now,
\begin{equation}\label{eq5}
 Z^{(d)}(T)=\exp\bigg[  N\beta V\Bigg( p^{(d)}_{\text{free}}\big(T,m+m_b\big) +  \frac{m^{d}}{ \alpha}\bigg(\frac{1}{d}\bigg)\bigg(\frac{d-1}{d}\bigg)^{d-1}\Bigg)\bigg],
\end{equation}
where \(\beta V\) is the spacetime volume and \(N\) is still the number of fermionic fields. Further, \(p^{(d)}_{\text{free}} \big(T,m\big)\) is the pressure of a free fermionic field theory at temperature \(T\) in \(d\) dimensions, which is given by, 
\begin{equation}\label{eq6}
  p^{(d)}_{\text{free}}(T,m)= d\int \frac{d^{d-1}p}{(2\pi)^{d-1}}\bigg[\frac{E_p}{2} + T\ln (1 + e^{-\beta E_p})\bigg]. 
\end{equation}
The partition function in \(3+1d\) is then
\begin{equation}
     Z=  e^{N\beta V \big(4J(T,m+m_b) +\frac{m^4}{4 \alpha}(\frac{3}{4})^3\big)}.
\end{equation}
Where in \(3+1d\)
\begin{equation}
    J(T,m) = \int d^3p \bigg[\frac{E_p}{2} + T\ln(1+ e^{-\beta E_p})\bigg],
\end{equation}
and \(E_p = \sqrt{\Vec{p^2}+(m+m_b)^2}\). Using dimensional regularization by perturbing around the dimension parameter \(d\), such that \(d= 4-2\epsilon\), \(J(T,m)\) evaluates to

\begin{equation}
    J(T,m)=-\frac{m^4}{64\pi^2}\left[\frac{1}{\epsilon}+\ln\left(\frac{\Bar{\mu}^2e^{3/2}}{m^2}\right)\right]-\frac{m^2T^2}{2\pi^2}\sum_{n=1}^{\infty}\frac{(-1)^n}{n^2}K_2\left(n\beta m\right),
\end{equation}
 where \(K_2(\cdot)\) denotes the modified Bessel function of the second kind.
In adding a renormalization scale \(\Bar{\mu}\), in the \(\overline{MS}\) scheme the pressure per fermionic component is
\footnote{
We note the pressure in the form of equation \eqref{eq10} is reminiscent of the usual GN-model where the the 2nd and 3rd terms represent a gas of quasiparticles of an effective mass \(m\).} 
\begin{equation}\label{eq10}
    p^{(4)}(T,m)=\frac{ (m-m_b)^4}{4\alpha}\bigg(\frac{3}{4}\bigg)^3-\frac{m^4}{16\pi^2}\left[\frac{1}{\epsilon}+\ln\left(\frac{\Bar{\mu}^2e^{3/2}}{m^2}\right)\right]-\frac{2m^2T^2}{\pi^2}\sum_{n=1}^{\infty}\frac{(-1)^n}{n^2}K_2\left(n\beta m\right). 
\end{equation}
Next, the pressure can then be non-perturbatively renormalized by absorbing \(\epsilon\) into the coupling such that
\begin{equation}
    \frac{m^4}{4\alpha}\bigg(\frac{3}{4}\bigg)^3- \frac{m^4}{16\pi^2}\frac{1}{\epsilon} = \frac{m^4}{4\alpha_R(\Bar{\mu})}
\end{equation}
where \(\alpha_R\) is the renormalized coupling. With this, the zero temperature piece of the pressure is
\begin{equation}
    p^{(4)}(T=0,m) =  \frac{m^4}{4\alpha_R(\bar{\mu})}-\frac{m^4}{16\pi^2}\ln\left(\frac{\Bar{\mu}^2e^{3/2}}{m^2}\right),
\end{equation}
and has a corresponding negative \(\beta\)-function of
\begin{equation}
\beta(\alpha_R)\equiv\frac{\partial \alpha_R(\Bar{\mu})}{\partial \ln(\Bar{\mu})} = -\frac{\alpha^2_R(\bar{\mu})}{2\pi^2}.
\end{equation}
Upon integration, there is a pole at some cut-off \(\Lambda_{c}\),
\begin{equation}
    \alpha_R(\Bar{\mu}) =\frac{4\pi^2}{\ln\left(\frac{\Bar{\mu}^2}{\Lambda^2_{c}}\right)}.
\end{equation}
We note that our result for \(\beta(\alpha_R)\) is \(-\beta(\lambda_R)\) where \(\beta(\lambda_R)\) is the \(\beta\)-function for a large-\(N\) scalar with quartic interactions given by \cite{Romatschke2022}. The renormalized pressure in terms of \(\alpha_R(\Bar{\mu})\) now reads 
\begin{align}\label{eq15}
    p^{(4)}(m,T)=&\frac{ -4 m^3 m_0+6 m^2 m_0^2+m_0^4-4 m m_0^3}{4\alpha}\bigg(\frac{3}{4}\bigg)^3 \\
    &-\frac{m^4}{16\pi^2}\ln\left(\frac{\Lambda_c^2 e^{3/2}}{m^2}\right)-\frac{2m^2T^2}{\pi^2}\sum_{n=1}^{\infty}\frac{(-1)^n}{n^2}K_2\left(n\beta m\right). \nonumber
\end{align}
However \(\frac{1}{\alpha}\) diverges as \(\epsilon\rightarrow 0\), therefore the remaining term which mixes bare and effective masses must be renormalized. To do this let
\begin{equation}
    \frac{m_b}{\alpha}\bigg(\frac{3}{4}\bigg)^3 = \frac{m_R(\Bar{\mu})}{\alpha_R(\Bar{\mu})}. 
\end{equation}
In demanding \(\frac{\partial p(m)}{\partial \ln (\Bar{\mu})}=0\) we have the expression

\begin{align}
    p(m,T)=&-m^3\frac{m_R(\Bar{\mu})}{\alpha_R(\Bar{\mu})}+\frac{6 m^2 m_0^2+m_0^4-4 m m_0^3}{4\alpha}\bigg(\frac{3}{4}\bigg)^3 \\
    &-\frac{m^4}{16\pi^2}\ln\left(\frac{\Lambda_c^2 e^{3/2}}{m^2}\right)-\frac{2m^2T^2}{\pi^2}\sum_{n=1}^{\infty}\frac{(-1)^n}{n^2}K_2\left(n\beta m\right)\nonumber. 
\end{align}

Now looking at higher powers of \(m_0\) we note
\begin{equation}
\begin{split}
    &\frac{m_b^2}{\alpha}=\bigg(\frac{m_R(\Bar{\mu})}{\alpha_R(\Bar{\mu})}\bigg)^2 \alpha
    = \text{Const}^2\frac{1}{\frac{1}{\alpha_R(\Bar{\mu})} + \frac{1}{4\pi^2\epsilon}}. 
\end{split}    
\end{equation}
Thus higher powers of the bare mass \(m_0\) are vanishing under our renormalization scheme as \(\epsilon \rightarrow 0\) and the renormalized pressure becomes
\begin{equation}
\begin{split}
    &p(m,T)=-m^3\frac{ m_R(\Bar{\mu})}{\alpha_R(\Bar{\mu})}-\frac{m^4}{16\pi^2}\ln\left(\frac{\Lambda_c^2 e^{3/2}}{m^2}\right)-\frac{2m^2T^2}{\pi^2}\sum_{n=1}^{\infty}\frac{(-1)^n}{n^2}K_2\left(n\beta m\right).
\end{split}   
\end{equation}
The condition \(\frac{\partial p_\text{ren}}{\partial\ln\Bar{\mu}}=0\) is fulfilled for
\begin{equation}
    m_R(\Bar{\mu}) = \frac{\text{const}}{\ln\big(\frac{\Bar{\mu}^2}{\Lambda^2}\big)}
\end{equation}
and we can define the constant $M\equiv \frac{m_R}{\alpha_R}$ as the renormalized bare mass. Finally, the renormalized pressure is given as,
\begin{equation}
      p(m,T)= -m^3 M-\frac{m^4}{16\pi^2}\ln\left(\frac{\Lambda_c^2 e^{3/2}}{m^2}\right)-\frac{2m^2T^2}{\pi^2}\sum_{n=1}^{\infty}\frac{(-1)^n}{n^2}K_2\left(n\beta m\right).
\end{equation}

\section{The $\mathcal{PT}$ Symmetric extension at Large-\(N\)}\label{section 3}
We now consider the \(\mathcal{PT}\)-symmetric version of equation \eqref{eq5} by analytically continuing $\alpha\rightarrow -g\pm i0^+$ such that
\begin{equation}
    \ln Z^{\mathcal{PT}} (g) = \frac{1}{2}\ln Z_{\text{Herm}}(\alpha \rightarrow{-g +i0^+}) + \frac{1}{2}\ln Z_{\text{Herm}}(\lambda \rightarrow{-g -i0^+}), 
\end{equation}
as conjectured by \cite{bender2022pt}. The analytically continued partition function is now
\begin{equation}\label{eq4}
 Z^{(d)}(T)=\exp\bigg[  N\beta V\Bigg( p^{(d)}_{\text{free}}\big(T,m\big) -  \frac{m^{d}}{ (g\pm i0^+)}\bigg(\frac{1}{d}\bigg)\bigg(\frac{d-1}{d}\bigg)^{d-1}\Bigg)\bigg],
\end{equation}
where in this expression we have let the bare mass \(m_0 =0\) for simplicity, provided we know how to reinsert the renormalized bare mass when needed from the previous section. Under dimensional regularization the pressure of the \(\mathcal{PT}\)-symmetric theory is given as
\begin{equation}
    p(m,T)=-\frac{m^4}{4g}\bigg(\frac{3}{4}\bigg)^3-\frac{m^4}{16\pi^2}\left[\frac{1}{\epsilon}+\ln\left(\frac{\Bar{\mu}^2e^{3/2}}{m^2}\right)\right]-\frac{2m^2T^2}{\pi^2}\sum_{n=1}^{\infty}\frac{(-1)^n}{n^2}K_2\left(n\beta m\right),
\end{equation}
 and is non-perturbatively renormalized through the coupling parameter by letting
\begin{equation}
    \frac{m^4}{4g_R(\Bar{\mu})}=\frac{m^4}{4g}\bigg(\frac{3}{4}\bigg)^3+\frac{m^4}{16\pi^2\epsilon},
\end{equation}
were \(g_R(\Bar{\mu})\) is the renormalized coupling. Now as a consequence of analytically continuing the coupling the \(\beta\)-function is positive and decaying into the IR,
\begin{equation}
\begin{split}
    &\beta(g_R) = \frac{g^2_R(\bar{\mu})}{2\pi^2}, \quad\text{and}\quad
     g_R(\Bar{\mu})=-\frac{4\pi^2}{\ln\left(\frac{\Bar{\mu}^2}{\Lambda^2_{c}}\right)}  \qquad \bar{\mu}<\Lambda_c,
\end{split}
\end{equation}
yielding a theory that is well-defined in the low momentum limit. In re-inserting the renormalized bare mass \(m_0^{'}\) and likewise re-inserting the analytic continuation of the coupling the pressure becomes
\begin{equation}
   p(m,T) =  -m^3 M -\frac{m^4}{16\pi^2}\ln\left(\frac{(\Lambda_c^2-i0^+)e^{3/2}}{m^2}\right)-\frac{2m^2T^2}{\pi^2}\sum_{n=1}^{\infty}\frac{(-1)^n}{n^2}K_2\left(n\beta m\right). 
\end{equation}
We note that the analytic continuation of the coupling gets removed via the coupling renormalization scheme. With this, the only physical parameter that remains in place of the coupling is the cut-off \(\Lambda\), which now carries the analytic continuation.
By taking the real part of the analytically continued partition function the renormalized pressure is given as
\begin{equation}\label{eq22}
     p(m,T)=-m^3 M-\frac{m^4}{16\pi^2}\ln\left(\frac{\Lambda_c^2 e^{3/2}}{m^2}\right)-\frac{2m^2T^2}{\pi^2}\sum_{n=1}^{\infty}\frac{(-1)^n}{n^2}K_2\left(n\beta m\right).
\end{equation}
\begin{figure}[h]
\centering
\includegraphics[width=.9\textwidth]{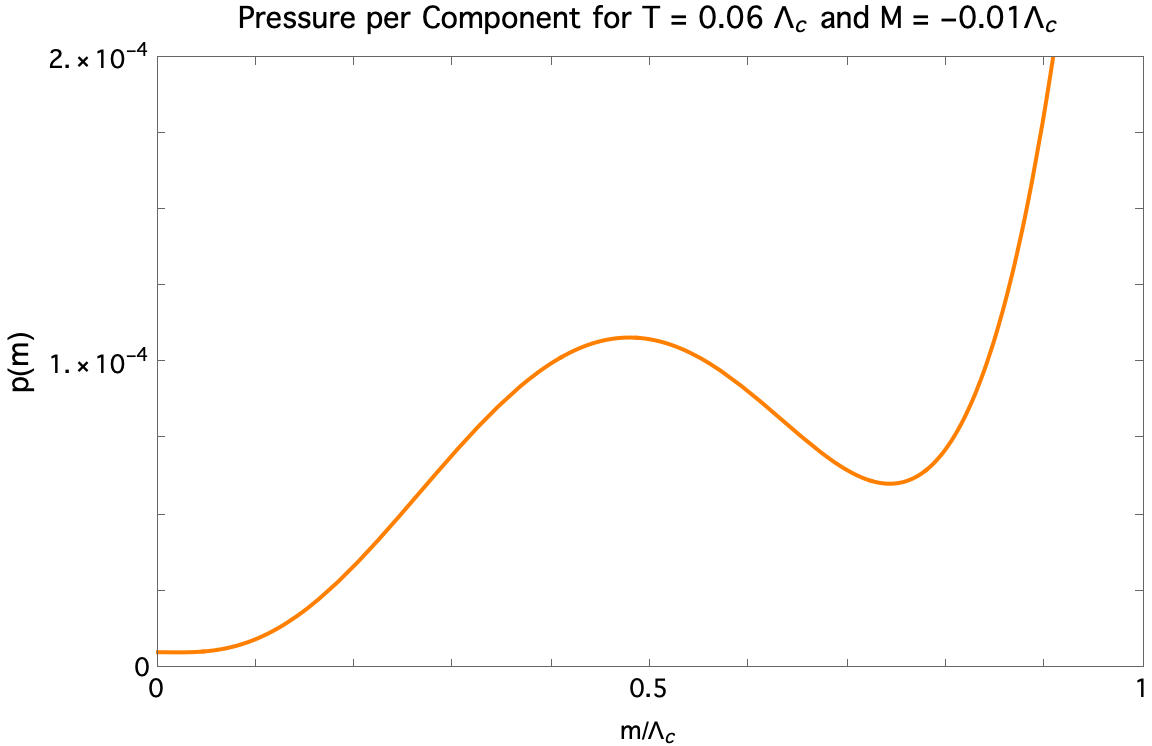}
\caption{\label{graph}The above figure gives the pressure per component of a Large-\(N\) theory of interacting fermions as a function of the effective mass, $m$. The plot is at a specific temperature \(T=.06\Lambda_c\) and mass scale \(M = -.01\Lambda_c\). We see that a non-zero solution to the saddle point equation (\ref{eq24}), around $m\approx 0.48\Lambda_c$, is thermodynamically favored.}
\label{fig: plot1}
\end{figure}
 Likewise the entropy density is calculated by the thermodynamic relation $s\equiv\frac{\partial p}{\partial T}$, to give
\begin{equation}
    s(m,T)= -\frac{2m^3}{\pi^2}\sum_{n=1}^{\infty}\frac{(-1)^nK_3(n\beta m)}{n},
\end{equation}
where $m$ is the solution to the gap equation.
Note that equation \eqref{eq15} is identical to equation \eqref{eq22}, thus the thermodynamics of the original theory, and the \(\mathcal{PT}\)-symmetric theory are identical. In this guise, we can view the coupling in the IR through the lens of the \(\mathcal{PT}\)-symmetric extension, and likewise the coupling in the UV can be viewed through the lens of the original theory. In this, we can see through the pole in running coupling and the theory becomes well-defined in both the IR and UV limits.

\section{Solving the Gap Equations of the \(\mathcal{PT}\)-Symmetric Model}

In both models, we have taken saddle point integrals around the auxiliary fields, and we have yet to consider gap equations of either theory. Generally, the partition function can be written  in the large $N$ limit as
\begin{equation}
    \lim_{N\gg 1}Z\left(\lambda\rightarrow -g+i0^+\right) = \sum_{i=1}^K e^{N\beta V p(m_i)},
\end{equation}
where $K$ enumerates the saddle points at each $m_i$ which minimizes the pressure. In taking the derivative of the dimensionally regulated pressure with respect to \(m\) (which is generated by the auxiliary field) the saddle point condition of the \(\mathcal{PT}\)-symmetric theory is given by
\begin{equation}\label{eq24}
    0=\frac{d p}{d m}=-3 M m^2-\frac{m^3 \left(1+\log \left(\frac{\Lambda_c ^2-i0^+}{m^2}\right)\right)}{4 \pi ^2}+\frac{2m^2 T}{\pi^2}\sum _{n=1}^{\infty } \frac{(-1)^n K_1(m n \beta )}{n}.
\end{equation}
\begin{figure}[h]
\centering
\includegraphics[width=.9\textwidth]{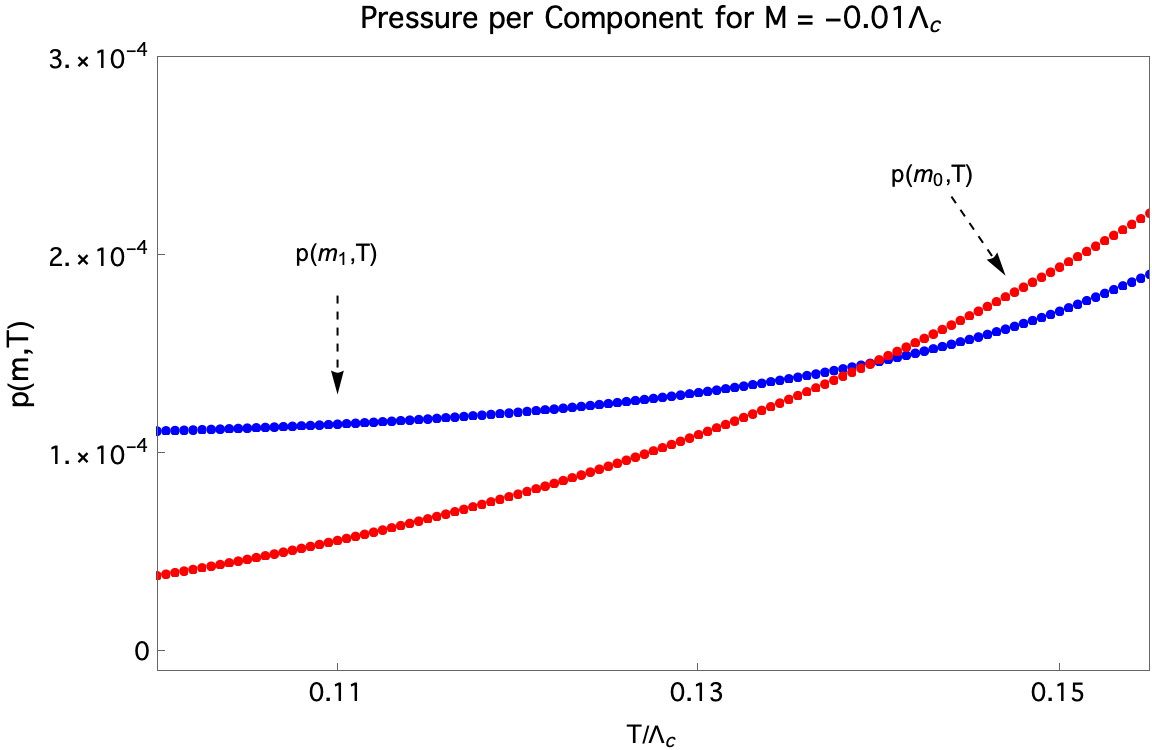}
\caption{\label{graph} This plot shows the $m_0=0$ solution to the gap equation becomes thermodynamically favored around \(T=0.139\Lambda_c\). }
\label{fig: plot2}
\end{figure}
 We first consider the analytic gap equations at zero temperature and zero bare mass \(T=M=0\). With this, the gap equation is given by
\begin{equation}
    m^3\log \left( \frac{\Lambda_c^2 e^1}{m^2} \right)=0,
\end{equation}
and has two solutions:
\begin{equation}\label{26}
    m_0(T=0)=0,\quad\text{and} \quad m_1(T=0)=\sqrt{e}\Lambda_c. 
\end{equation}
We note that \(m_1\) depends on the scale \(\Lambda_c\) and \(\Lambda_c\) depends on the coupling \(\alpha_R\). Thus, \(m_1\) is explicitly related to the GN-model, and must be physical. Now the partition function evaluates to
\begin{equation}
    \lim_{N\gg 1}Z\left(\lambda=-g,T=0\right) = e^{-N\beta V \Lambda_c^4 e^2/32\pi^2} + 1.
\end{equation}
We note that zero temperature gap equations of our fermionic theory are equivalent for not only both Hermitian and \(\mathcal{PT}\)-symmetric theories, but they also match the zero temperature gap solutions of \(\phi\)-four scalar theory in \(3+1d\) given by \cite{Romatschke2022}. 

Next, we consider re-inserting the bare mass term. The zero-temperature solutions to our gap equations are now given by

\begin{equation}
    m = \frac{6 \pi ^2 M}{W\left(-6 \pi ^2 \sqrt{\frac{M^2}{e\Lambda_c ^2}}\right)}\quad \& \quad m= \frac{6 \pi ^2 M}{W\left(6 \pi ^2 \sqrt{\frac{M^2}{e\Lambda_c ^2}}\right)}.
\end{equation}

\begin{figure}[h]
\centering
\includegraphics[width=.9\textwidth]{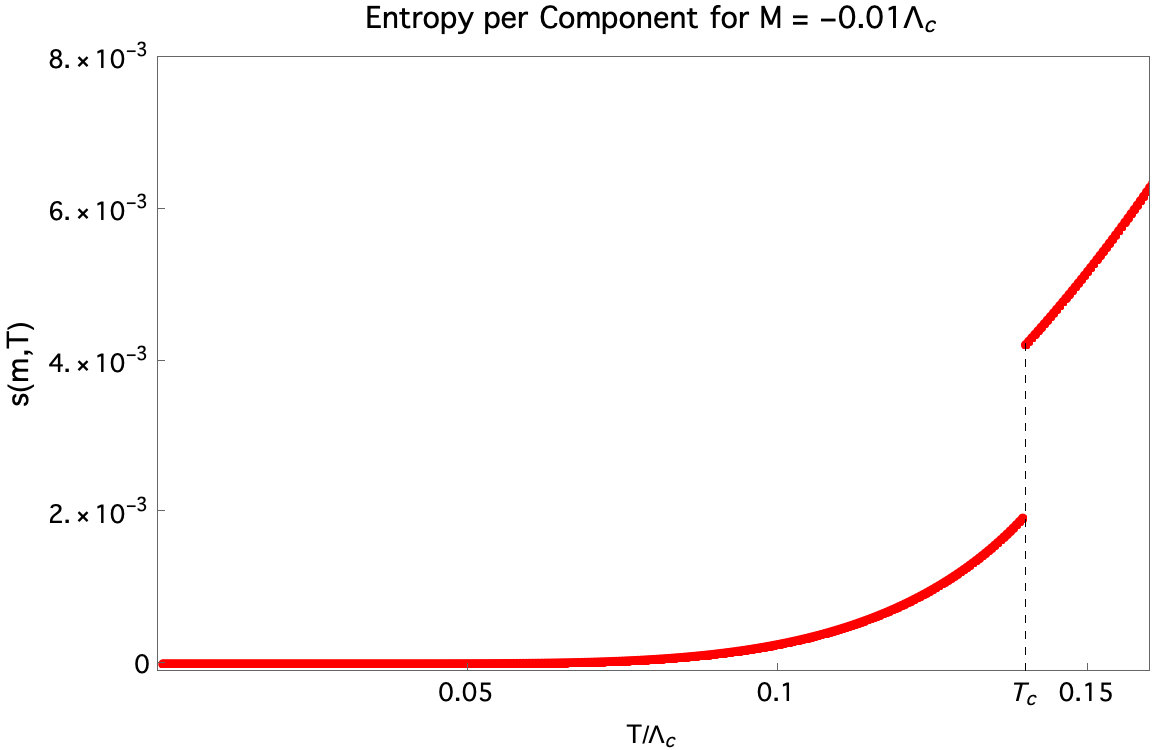}
\caption{\label{graph} The entropy density has a discontinuity at $T=T_c$ signaling a first-order phase transition.}
\label{fig: plot3}
\end{figure}

Where \(W(\cdot)\) is the Lambert \(W\)-functuion. In letting \(M=-.01\Lambda_c\) we find three solutions to the zero temperature gap equation at \(m_0=0\), \(m_1=0.47945\Lambda_c\), and \(m_2=0.74319\Lambda_c\), shown in figure (\ref{fig: plot1}). At finite temperatures, the in-medium mass solutions to (\ref{eq24}) can be solved numerically. The pressure can then be evaluated at these solutions to determine which ones are thermodynamically favored and which ones are meta-stable. With this, the pressure is greatest for the temperature-dependent solution $m_1(T)$ up until \(T_c\approx 0.139\Lambda_c\) when the \(m_0=0\) solution becomes thermodynamically favored as shown in figure \ref{fig: plot2}. When we evaluate the entropy density we find a first-order phase transition which can be seen by a discontinuity at $T_c$ in figure \ref{fig: plot3}.

\section{Conclusion}

In this paper, we constructed two theories of fermionic interactions which share the same physical observables. The overlap of these theories gives a combined theory which is UV and IR complete, renormalizable, and fully solvable for all couplings. This theory is shown to have a stable, meta-stable, and unstable phase which are separated by a first-order phase transition. Further, our theory contains a meta-stable state in the pressure and a first-order phase transition around \(T=0.139\Lambda_c\). Generally, we found that at large \(N\) the gap equations and the pressures of our theories were invariant to sign changes in the coupling, as anticipated. 

This work motivates several paths for future work. As mentioned in the introduction, the NJL has a rich history, motivating the study of a model with marginal pseudo-scalar four-fermi interactions of the form $(\Bar{\psi}\gamma_5\psi)^2$. As the NJL which model contains continuous chiral symmetry instead of only discrete, the model provides (pseudo) Goldstone bosons to describe pions for example. If our model could be generalized to consider a pseudo-scalar four fermion channel perhaps one could move beyond the drawbacks of the standard NJL-model, that being its nonrenormalizability and a a cutoff of about 700 MeV. 

Further, we conjecture that similar methods can be used to add deformations of relevant fermionic operators to the marginal theory in \(3+1d\). Likewise, one could extend our theory, (or a relevant version of our theory) with a finite chemical potential, and look for a tricritical point that is phenomenologically similar to QCD.
Working in lower dimensions we also speculate that the benign nature of the poles found at large-\(N\) will hold in \(1+1d\). One could show this explicitly by looking at theories with changes in signs of the couplings. If so, we speculate such results could be used to experimentally see past critical scales associated with \(1+1d\) relativistic fermion interactions and spin chains. 
 Further, novel changes in the sign of the coupling at large-\(N\) can be explored in making UV and IR complete large-\(N\) theories for QED and electroweak theory, and likewise large-\(N\) toy models for QCD. This and many other papers have shown how the coupling independent expansion parameter of \(\frac{1}{N}\) can lead to fully solvable theories for all couplings, making the general topic of large-\(N\) theories a fascinating and fruitful subject with ample room for exploration.
\section{Acknowledgements}
We thank Paul Romatschke, Scott Lawrence, Ryan Weller, and Marcus Pinto for the abundance of insightful feedback and discussion. This work was supported by the DOE, award number DE- SC0017905.

\end{spacing}

\bibliographystyle{plain}
\bibliography{bibliography.bib}

\begin{thebibliography}{10}

\bibitem{alexandre2018spontaneous}
Jean Alexandre, John Ellis, Peter Millington, and Dries Seynaeve.
\newblock
  \href{https://journals.aps.org/prd/abstract/10.1103/PhysRevD.98.045001}{Spontaneous
  symmetry breaking and the Goldstone theorem in non-Hermitian field theories}.
\newblock {\em Physical Review D}, 98(4):045001, 2018.

\bibitem{banerjee1997explicit}
R~Banerjee and EC~Marino.
\newblock
  \href{https://journals.aps.org/prd/abstract/10.1103/PhysRevD.56.3763}{Explicit
  bosonization of the massive Thirring model in 3+ 1 dimensions}.
\newblock {\em Physical Review D}, 56(6):3763, 1997.

\bibitem{bender1998real}
Carl~M Bender and Stefan Boettcher.
\newblock
  \href{https://journals.aps.org/prl/abstract/10.1103/PhysRevLett.80.5243}{Real
  spectra in non-Hermitian Hamiltonians having P T symmetry}.
\newblock {\em Physical review letters}, 80(24):5243, 1998.

\bibitem{bender1999pt}
Carl~M Bender, Stefan Boettcher, and Peter~N Meisinger.
\newblock
  \href{https://aip.scitation.org/doi/abs/10.1063/1.532860?casa_token=JQZ1jlLnOdwAAAAA:oLi0Tvf5yNqwSUDxx0Cl8NYWfxebwUq413mhQPzP8P279u4BjbfHZlq5RSGaZzF68T6RrI0Mfz8}{PT-symmetric
  quantum mechanics}.
\newblock {\em Journal of Mathematical Physics}, 40(5):2201--2229, 1999.

\bibitem{BenderFermi}
Alireza Beygi, S.~P. Klevansky, and Carl~M. Bender.
\newblock
  \href{https://journals.aps.org/pra/abstract/10.1103/PhysRevA.99.062117}{Relativistic
  $\mathcal{PT}$-symmetric fermionic theories in $1+1$ and $3+1$ dimensions}.
\newblock {\em Phys. Rev. A}, 99:062117, Jun 2019.

\bibitem{chernodub2021spontaneous}
Maxim~N Chernodub, Alberto Cortijo, and Marco Ruggieri.
\newblock
  \href{https://journals.aps.org/prd/abstract/10.1103/PhysRevD.104.056023}{Spontaneous
  non-hermiticity in the Nambu--Jona-Lasinio model}.
\newblock {\em Physical Review D}, 104(5):056023, 2021.

\bibitem{diab2016and}
Kenan Diab, Lin Fei, Simone Giombi, Igor~R Klebanov, and Grigory Tarnopolsky.
\newblock
  \href{https://iopscience.iop.org/article/10.1088/1751-8113/49/40/405402/meta?casa_token=pVodmwk1Au4AAAAA:u-VwwVMdUwDQ_YJItqiKb_SCd3Dta-UEuUU8p9CuCEtEVf3jYSIQmtbqUAGGBoKc3NwDW43_KvTOuZp8uJN6}{On
  and in the Gross--Neveu and O (N) models}.
\newblock {\em Journal of Physics A: Mathematical and Theoretical},
  49(40):405402, 2016.

\bibitem{fring2020pseudo}
Andreas Fring and Takanobu Taira.
\newblock
  \href{https://journals.aps.org/prd/abstract/10.1103/PhysRevD.101.045014}{Pseudo-Hermitian
  approach to Goldstone’s theorem in non-Abelian non-Hermitian quantum field
  theories}.
\newblock {\em Physical Review D}, 101(4):045014, 2020.

\bibitem{fring2020goldstone}
Andreas Fring and Takanobu Taira.
\newblock
  \href{https://www.sciencedirect.com/science/article/pii/S0550321319303207}{Goldstone
  bosons in different PT-regimes of non-Hermitian scalar quantum field
  theories}.
\newblock {\em Nuclear Physics B}, 950:114834, 2020.

\bibitem{fring2022massive}
Andreas Fring and Takanobu Taira.
\newblock
  \href{https://link.springer.com/article/10.1140/epjp/s13360-022-02889-z}{Massive
  gauge particles versus Goldstone bosons in non-Hermitian non-Abelian gauge
  theory}.
\newblock {\em The European Physical Journal Plus}, 137(6):1--15, 2022.

\bibitem{grable2022theremal}
Seth Grable.
\newblock
  \href{https://link.springer.com/article/10.1007/JHEP10(2022)133}{Interacting
  CFTs for all couplings: thermal versus entanglement entropy at large N.}
\newblock {\em J. High Energ. Phys.}, 133, 2022.

\bibitem{gracey1990three}
JA~Gracey.
\newblock
  \href{https://www.sciencedirect.com/science/article/pii/055032139090186H}{Three-loop
  calculations in the O (N) Gross-Neveu model}.
\newblock {\em Nuclear Physics B}, 341(2):403--418, 1990.

\bibitem{gross1974dynamical}
David~J Gross and Andre Neveu.
\newblock
  \href{https://journals.aps.org/prd/abstract/10.1103/PhysRevD.10.3235}{Dynamical
  symmetry breaking in asymptotically free field theories}.
\newblock {\em Physical Review D}, 10(10):3235, 1974.

\bibitem{laine2016basics}
Mikko Laine and Aleksi Vuorinen.
\newblock
  \href{https://link.springer.com/content/pdf/10.1007/978-3-319-31933-9.pdf}{Basics
  of thermal field theory}.
\newblock {\em Lect. Notes Phys}, 925(1):1701--01554, 2016.

\bibitem{mannheim2019goldstone}
Philip~D Mannheim.
\newblock
  \href{https://journals.aps.org/prd/abstract/10.1103/PhysRevD.99.045006}{Goldstone
  bosons and the Englert-Brout-Higgs mechanism in non-Hermitian theories}.
\newblock {\em Physical Review D}, 99(4):045006, 2019.

\bibitem{Mavromatos:2021hpe}
N.~E. Mavromatos, Sarben Sarkar, and A.~Soto.
\newblock
  {\href{https://journals.aps.org/prd/pdf/10.1103/PhysRevD.106.015009}{PT
  symmetric fermionic field theories with axions: Renormalization and dynamical
  mass generation}}.
\newblock {\em Phys. Rev. D}, 106(1):015009, 2022.

\bibitem{mavromatos2023schwinger}
NE~Mavromatos, Sarben Sarkar, and A~Soto.
\newblock
  \href{https://www.sciencedirect.com/science/article/pii/S0550321322003996}{Schwinger-Dyson
  equations and mass generation for an axion theory with a PT symmetric Yukawa
  fermion interaction}.
\newblock {\em Nuclear Physics B}, 986:116048, 2023.

\bibitem{moshe2003quantum}
Moshe Moshe and Jean Zinn-Justin.
\newblock
  \href{https://www.sciencedirect.com/science/article/pii/S0370157303002631}{Quantum
  field theory in the large N limit: A Review}.
\newblock {\em Physics Reports}, 385(3-6):69--228, 2003.

\bibitem{parisi1975theory}
Giorgio Parisi.
\newblock
  \href{https://www.sciencedirect.com/science/article/pii/0550321375906240}{The
  theory of non-renormalizable interactions: The large N expansion}.
\newblock {\em Nuclear Physics B}, 100(2):368--388, 1975.

\bibitem{peskin2018introduction}
Michael~E Peskin.
\newblock {\em
  \href{https://www.taylorfrancis.com/books/mono/10.1201/9780429503559/introduction-quantum-field-theory-michael-peskin}{An
  introduction to quantum field theory}}.
\newblock CRC press, 2018.

\bibitem{pinto2020three}
Marcus~Benghi Pinto.
\newblock
  \href{https://journals.aps.org/prd/pdf/10.1103/PhysRevD.102.065005}{Three-dimensional
  Yukawa models and CFTs at strong and weak couplings}.
\newblock {\em Physical Review D}, 102(6):065005, 2020.

\bibitem{Romatschke2022}
Paul Romatschke.
\newblock \href{https://arxiv.org/abs/2211.15683}{A solvable quantum field
  theory with asymptotic freedom in 3+1 dimensions}.
\newblock {\em arXiv:2211.15683 [hep-th]}.

\bibitem{romatschke2019finite}
Paul Romatschke.
\newblock
  \href{https://journals.aps.org/prl/abstract/10.1103/PhysRevLett.122.231603}{Finite-temperature
  conformal field theory results for all couplings: O (N) model in 2+ 1
  dimensions}.
\newblock {\em Physical Review Letters}, 122(23):231603, 2019.

\bibitem{romatschke2021shear}
Paul Romatschke.
\newblock
  \href{https://journals.aps.org/prl/pdf/10.1103/PhysRevLett.127.111603}{Shear
  Viscosity at Infinite Coupling: A Field Theory Calculation}.
\newblock {\em Physical Review Letters}, 127(11):111603, 2021.

\bibitem{romatschke2019thermal}
Paul Romatschke and Saga S{\"a}ppi.
\newblock
  \href{https://journals.aps.org/prd/pdf/10.1103/PhysRevD.100.073009}{Thermal
  free energy of large N f QED in 2+ 1 dimensions from weak to strong
  coupling}.
\newblock {\em Physical Review D}, 100(7):073009, 2019.

\bibitem{Weiner:2022vwa}
Max Weiner and Paul Romatschke.
\newblock {\href{https://arxiv.org/abs/2208.10502}{Determining all
  thermodynamic transport coefficients for an interacting large N quantum field
  theory}}.
\newblock 8 2022.

\bibitem{bender2022pt}
Sarben~Sarkar Wen-Yuan~Ai, Carl M.~Bender.
\newblock \href{https://arxiv.org/abs/2209.07897}{PT-symmetric \(-g\phi^4\)
  Theory}.
\newblock {\em arXiv:2209.07897 [hep-th]}.

\end{thebibliography}
\end{document}